\documentclass[twocolumn,showpacs,preprintnumbers,prl,floatfix]{revtex4-1}
\usepackage{amssymb,amsmath,amsfonts}
\usepackage{graphics,dcolumn,bm,epic,eepic,float}
\usepackage{graphicx} 
\usepackage{epstopdf} 
\usepackage{bm} 
\usepackage{subfigure}
\usepackage{color} 

\begin{document}
\title{Understanding mobility in a social petri dish}

\author{Michael Szell$^{1}$, Roberta Sinatra$^{2,3}$, Giovanni Petri$^{4,5}$, Stefan Thurner$^{1,6,7}$, Vito Latora$^{2,3}$}

\affiliation{%
  $^1$~Section for Science of Complex Systems, Medical University of
  Vienna, Spitalgasse 23, 1090 Vienna, Austria} \affiliation{%
  $^2$~Dipartimento di Fisica e Astronomia, Universit\`a di Catania
  and INFN, Via S. Sofia, 64, 95123 Catania, Italy} \affiliation{%
  $^3$~Laboratorio sui Sistemi Complessi, Scuola Superiore di Catania,
  Via San Nullo 5/i, 95123 Catania, Italy} \affiliation{%
  $^4$~Centre for Transport Studies, Department of Civil and
  Environmental Engineering, Imperial College London, London SW7 2AZ,
  UK} \affiliation{%
  $^5$~Complexity and Networks group, Imperial College London, London
  SW7 2AZ, UK} \affiliation{%
  $^6$~Santa Fe Institute, Santa Fe, NM 87501, USA} \affiliation{%
  $^7$~IIASA, Schlossplatz 1, 2361 Laxenburg, Austria}
\date{\today}

\begin{abstract}  
Despite the recent availability of large data sets on human movements, a full understanding of the rules governing motion within social systems is still missing, due to incomplete information on the socio-economic factors and to often limited spatio-temporal resolutions.
Here we study an entire society of individuals, the players of an online-game, with complete information on their movements in a network-shaped universe and on their social and economic interactions.
Such a ``socio-economic laboratory'' allows to unveil the intricate interplay of spatial constraints, social and economic factors, and patterns of mobility.
We find that the motion of individuals is not only constrained by physical distances, but also strongly shaped by the presence of socio-economic areas. These regions can be recovered perfectly by community detection methods solely based on the measured human dynamics.
Moreover, we uncover that long-term memory in the time-order of visited locations is the essential ingredient for modeling the trajectories.

\end{abstract}
\maketitle

Understanding the statistical patterns of human mobility, predicting
trajectories and uncovering the mechanisms behind human movements
\cite{barthelemy2010sn} is a considerable challenge with important
practical applications to traffic management \cite{guimera2005wan,
  helbing2001trs}, planning of urban spaces \cite{makse1995mug,
  roth2010cpc}, epidemics \cite{pastor2001ess, colizza2006rat,
  hufnagel2004fce,balcan2009mmn}, information spreading \cite{miritello2011dss,
  onnela2007sts}, and geo-marketing \cite{quercia2010rse, jensen2006npr}.
In the last years, advanced digital technologies have provided huge
amounts of data on human activities, allowing to extract 
information on human movements. 
For instance, observations of banknote
circulation \cite{brockmann2006slh, thiemann2010sbs}, mobile phone
records \cite{gonzalez2008uih}, online location-based social networks
\cite{scellato2011ssp, scellato2011nps}, GPS location data of vehicles
\cite{bazzani2010slu}, or radio frequency identification  traces
\cite{barthelemy2010sn, cattuto2010dpi, roth2010cpc}, have all been
used as proxies for human movements.
These studies have provided valuable insights into several aspects of
human mobility, uncovering distinct features of human travel behaviour
such as scaling laws \cite{brockmann2006slh, song2010msp},
predictability of trajectories \cite{song2010lph}, and impact of
motion on disease spreading \cite{hufnagel2004fce, belik2011nhm,
  colizza2006rat,balcan2009mmn}.
\begin{figure}[t!]              
    \begin{center}
        \includegraphics[width=0.412\textwidth]{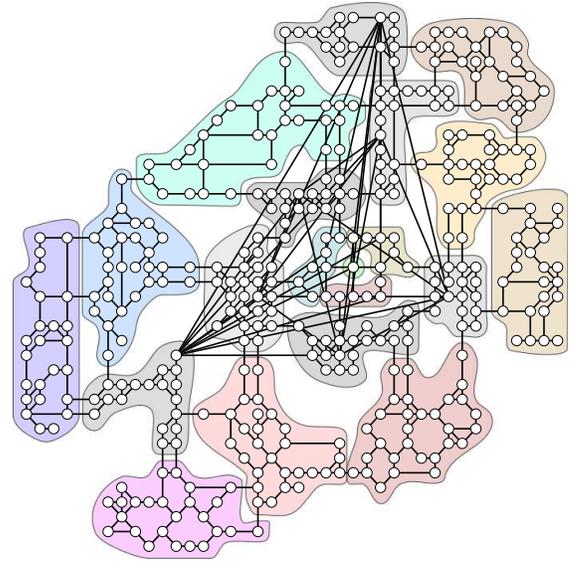}    
    \end{center}
    \caption{\footnotesize {\bf The universe map of the massive multiplayer
        online game \emph{Pardus}.} The universe of Pardus can be
      represented as a network \cite{boccaletti2006cns} with $N=400$ nodes, called sectors (playing the role of cities), and
      $K=1160$ links. Sectors are organized into 20
      different regions, called clusters, shown in the figure as
      different colour-shaded areas. There is no explicit set of goals in
      the game. Players are free to interact in a number of ways to
      e.g.~increase their virtual wealth or status. Players move
      between sectors to interact with other players, e.g.~to trade,
      attack, wage war, or to explore the virtual
      world.\label{fig:universe}}
\end{figure}
However, from a comparative analysis
of the different works it emerges clearly that a ``unified theory'' of
human mobility is still outstanding, since results, even on some very
basic features of the motion, often appear to be contrasting
\cite{barthelemy2010sn}. One example is the measured distribution of
human trip lengths in various types of transportation: some studies agree that mobility is generally
characterized by fat-tailed distributions of trip lengths
\cite{brockmann2006slh, song2010msp}, while others report exponential
or binomial forms \cite{bazzani2010slu, roth2010cpc,
  barthelemy2010sn}.
The discrepancies arise due to the different mobility data sets used,
where mobility is indirectly inferred from some specific human
activity in a particular context. For instance, mobile phone records
typically provide location information only when a person uses the
phone \cite{song2010msp}, while radio frequency identification traces
like the ones of Oyster cards in the London subway \cite{roth2010cpc} only log
movements based on public transportation systems. Analyses of these
data sets can then result in a possibly biased view of the underlying
mobility processes.
Furthermore, most of the analyzed data sets have poor information on
how socio-economic factors influence human mobility patterns.  More
generally, the lack of an all-encompassing record set with positional
raw data, including complete information on the socio-economic context
and on the behaviour of all members of a human society, has so far
limited the possibilities for a comprehensive exploration of human
mobility.

Here, we address the issue of mobility from a novel point of view by
analyzing, with unprecedented precision, the movements of a large number
of individuals, the players 
of a self-developed massive multiplayer online game (MMOG). Such
online platforms provide a fascinating new way of observing hundreds
of thousands of interacting individuals who are simultaneously engaged in
social and economic activities.  The potential of online worlds as
large-scale ``socio-economic laboratories'' has been demonstrated in a
number of previous studies \cite{szell2010msd, szell2010mol,
  castronova2006orv, bainbridge2007srp}. For the MMOG
at hand \cite{pardus}, we have access to practically all actions \cite{thurner2011egc},
including movements, accumulated over several years. This MMOG can therefore be considered as a ``socio-economic petri dish'' to study
mobility in a completely controlled way. We can in fact observe the
long-time evolution of a social system at the scale of an entire human
society, having a perfect knowledge of all the spatio-temporal
and socio-economic details. In contrast to traditional studies in
social science which are typically biased by well-known
``interviewer-effects'', in MMOGs the socio-economic measurements are objective and unobtrusive,
since subjects are not consciously aware of being observed.

Using positional data of the players in the game universe
\footnote{Whenever we address the position or the movement of `a
  player' in the game universe, this is meant as a short form for
  referring to the virtual avatar which is uniquely associated to and
  controlled by the player. This abbreviation is consistent with the
  tendency of players to identify themselves with their avatars.}, in combination with
other socio-economic information from the game, we uncover various
fundamental features of mobility, and we provide a complete
description of the mechanisms causing the observed anomalous
diffusion. Two are the main results of our work. 
First, we find the
emergence of different spatial scales, due to the strong tendency of
the players to limit their economic activities to some specific areas
over long time periods and to avoid crossing the borders between
different areas. Making use of this observation, 
  we propose an efficient method to identify
socio-economic regions by means of community detection algorithms
based solely on the measured movement dynamics.
Our second result unveils the driving
mechanism behind the movement patterns of players: 
Locations are visited in a specific order, leading to strong 
long-term memory effects which are essential to understand and 
reproduce the observed trajectories. 
Finally, we provide large-scale evidence that neglecting either of
these spatial or temporal constraints may obstruct the possibility of 
understanding the processes behind human mobility.

\subsection{A social arena: the online game Pardus}
\emph{Pardus} is a massive multiplayer online game running
since 2004, with a worldwide player base of more than 350,000
individuals. It is an open-ended game whose players live in a virtual,
futuristic universe and interact with each other in a multitude of ways.
The topology of the universe can be represented as a network with $400$ nodes, called
\emph{sectors}, embedded in a two-dimensional space, the so-called
\emph{universe map} shown in Fig.~\ref{fig:universe}.  Each sector is like a
city where players can have social relations (establish new
friendships, make enemies and wage wars), and entertain economic
activities (trade and production of commodities). Typically, sectors
adjacent on the universe map, as well as a few far-apart sectors,
are interconnected by links which allow players to move from sector to
sector. 
At any point in time, each sector is usually attended by a large number of players.
The network is sparse and, similarly to other spatial
networks, is not a small world. It has a characteristic path length
$L=11.89$ and a diameter $d_{\mathrm{max}}=27$, which means that, on average, players
have to move through a non-negligible number of sectors to traverse
the universe. See \cite{SI} Section S3 and \cite{SI} Table~I for a detailed
characterization of the universe network structure.

The sectors have been originally organized by the developers of the
game into 20 different \emph{clusters}, which are perceived by the players as different
political or socio-economic regions such as countries. For example, a player who is member of a political faction in the game is provided some game-relevant protection in all clusters which are controlled by the faction, and has the opportunity of social promotion when accomplishing certain tasks within these clusters. Each cluster is shown in Fig.~\ref{fig:universe} with a different
background colour. All clusters contain about 20 sectors each, with the
exception of the central cluster, consisting of just one sector, and
its surrounding three clusters having only 6-7 sectors. Sectors
belonging to the same cluster are geographically close on the map,
meaning that the distance between any two sectors in the same cluster
is small, with an average distance around 3. Players typically have a ``home cluster'' where they focus their
socio-economic activities over long time periods. Occasionally, they also move to sectors belonging to other clusters in order to explore the
universe, to relocate their home (migrate), or during extreme game events
such as wars.

In Pardus, players are free to pursue whichever role they like to
take. Many of them focus on expanding their social relations
or political influence, some play the role of ``scientists'' exploring
the universe, while others choose their main goal in trade and
optimizing the amount of virtual money earned \cite{szell2010msd}.
The large variety of complex socio-economic behaviours emerging in this online society, results in high heterogeneity in the mobility
patterns, such as observed in real human motion.  However,
differently from other empirical studies on human movements, mobility
in Pardus can be investigated in a controlled way, since complete
information on actions of players is available \cite{szell2010msd,
  szell2010mol}. In this article we consider a data set consisting of movements in the network universe of all players who were active over a period of 1,000 days, as well as of socio-economic information about their environment.
This 
opens the possibility of investigating motion in relation to other
 social and economic factors. 
 Note that we do not have to address the common issues of relying on incomplete data, on data that are only a proxy of mobility, or on data that are aggregates of different types of transportation \cite{balcan2009mmn}. 
 See \cite{SI} Section S2 for more details on the data set.

\begin{figure}[t!]               
    \begin{center}
        \includegraphics{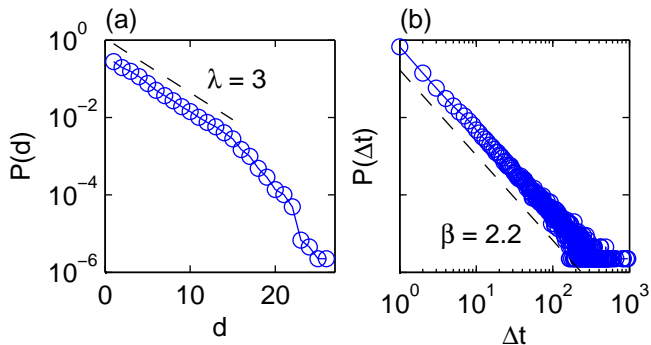}   
    \end{center}
    \caption{\footnotesize {\bf Distribution of jump distances and of
        waiting times.}  To each player a time series consisting of
      the sector positions over 1000 days is associated. A \emph{jump}
      is said to occur when the sector position in the time series
      changes from one day to the following.  The length $d$ of a jump
      is measured in terms of graph distance and can take an integer value
      between 1 and $d_{\mathrm{max}} = 27$, the diameter of the
      network. (a) The probability distribution of jump distances is
      reported in a semi-log plot. For $d\leq15$, the distribution follows an exponential
      $P(d) \sim e^{-\frac{d}{\lambda}}$ with a characteristic length
      $\lambda \approx 3$.  Players can also remain in the same sector for
      more days, without moving to other sectors.  We define as
      waiting time $\Delta t$ the number of consecutive days a player
      spends in only one sector. (b) We show the probability
      distribution of waiting times $\Delta t$ in a log-log plot, which is well fitted
      by a power-law $P(\Delta t) \sim \Delta t^{-\beta}$, with
      $\beta \approx 2.2$. \label{fig:distributions}}
  \end{figure}

\subsection{Basic features of the motion}

The position of each player in the universe, namely the ID number of the
sector where the player is currently situated, is logged once a day. 
In this way the motion of each player becomes a time 
series of 1,000 sector positions. A \emph{jump} occurs when a player's sector 
position changes from one day to the following.
 The associated length $d$ of a jump is
measured in terms of graph distance, an integer value between 1
and $d_{\mathrm{max}} = 27$. The
probability distribution of jump distances, computed for all players
over the whole observation period, is reported in
Figure~\ref{fig:distributions}~(a).  For $d \leq 15$, the distribution
is well-fitted by an exponential: 
\begin{equation}
P(d) \sim e^{-\frac{d}{\lambda}},
\label{eq:distancedistribution}
\end{equation}
with a characteristic jump length $\lambda \approx 3$. The existence of
a typical travel distance, as also
recently found in other mobility data \cite{roth2010cpc,bazzani2010slu}, 
is related to the use of a single
transportation mode in \emph{Pardus} \cite{koelbl2003energy}.
This allows to disentangle the intrinsic heterogeneity of the players
from the effects due to the presence of different means of
transportation \cite{balcan2009mmn}, 
which might be the cause of the scale-free distributions found in
mobile phone or other mobility data sets \cite{brockmann2006slh,
  gonzalez2008uih}. It has in fact been suggested that power laws in
distance distributions of movement data may emerge from the
coexistence of different scales
\cite{barthelemy2010sn,han2011osl}.

In some cases, players stay in the same sector for a number of
consecutive days.  For instance, 11 of the 1458 considered players, although
being active in the game, never jump within the entire observation
period. On average, a player does not change
sector in approximately $75\%$ of the days. To better characterize 
the motion, we computed the waiting times 
$\Delta t$ (measured in terms of number of days) 
between all pairs of consecutive jumps, over all players. 
The distribution of these waiting times, shown in 
Fig.~\ref{fig:distributions}~(b) follows a power-law distribution: 
\begin{equation}
P(\Delta t) \sim \Delta t^{-\beta} 
\label{eq:waitingtimedistribution}
\end{equation}
with an exponent $\beta \approx 2.2$, in agreement with other
recent measurements on human dynamics \cite{barabasi2005obh}. In
addition, we found that the average waiting times of 
individual players are distributed as a power-law (see \cite{SI} Fig.~2). This implies a strong 
heterogeneity in the motion of different players, which is 
related to the heterogeneity in their general
activity (see \cite{SI} Section S1 and \cite{SI} Fig.~1).

\begin{figure}[t!]            
    \begin{center}
        \includegraphics[width=0.33\textwidth]{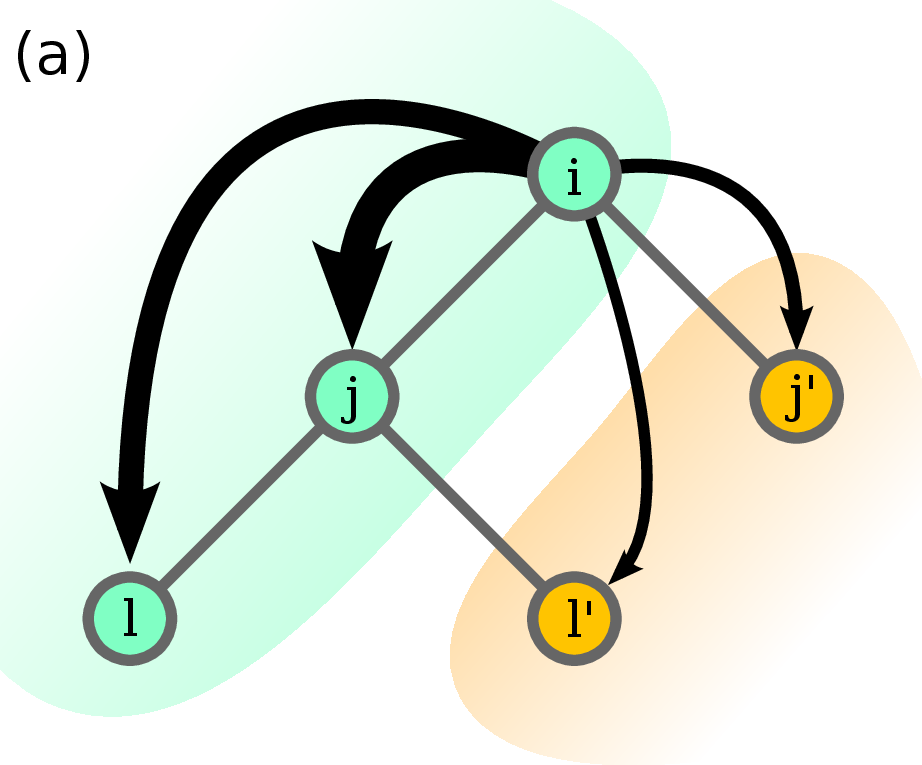}   
        \includegraphics{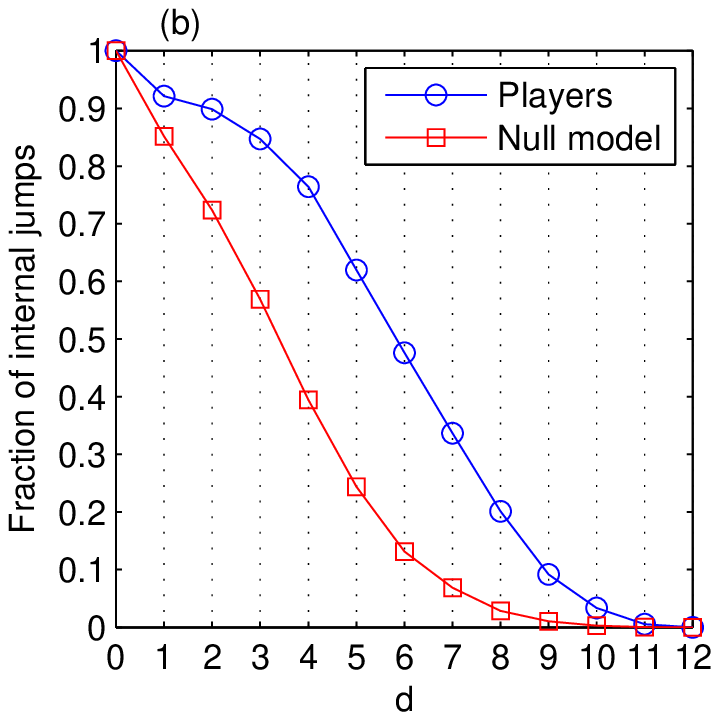}    
    \end{center}
    \caption{\footnotesize {\bf Influence of socio-economic clusters
        on mobility}. (a) Sketch of jump patterns from a sector $i$ to
      sectors within the same cluster, $j$ and $l$, and to sectors in
      a different cluster, $j'$, $l'$. Although sectors $j'$ and $l'$
      have the same graph distance from sector $i$ as sectors $j$ and
      $l$ respectively, transitions across cluster border have smaller
      probabilities. (b) Quantitative evidence of the tendency of
      players to avoid crossing borders.  Red squares show the
      null model, i.e. the fraction of all pairs of sectors at a given
      distance $d$ being in the same cluster.  Blue circles show the
      fraction of measured jumps leading into the same cluster, per
      distance.  Coincidence of the two curves would indicate that
      clusters have no effect on mobility. Clearly this is not the
      case -- there is a strong tendency of players to avoid crossing
      the borders between clusters.\label{fig:banana}}
  \end{figure}

\subsection{Mobility reveals socio-economic clusters} 

\begin{figure*}[t!]            
    \begin{center}
    \begin{tabular}{ccc}
      \hspace{-0.7cm}\includegraphics{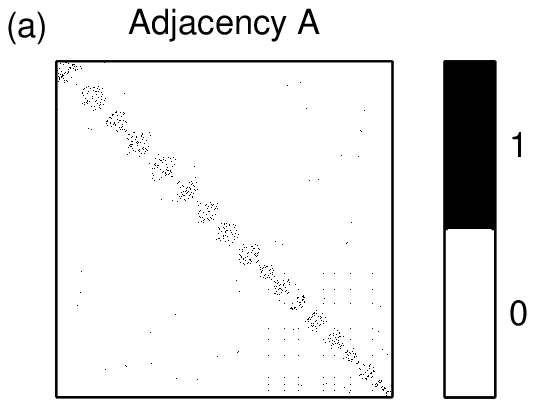} & \hspace{-0.7cm}\includegraphics{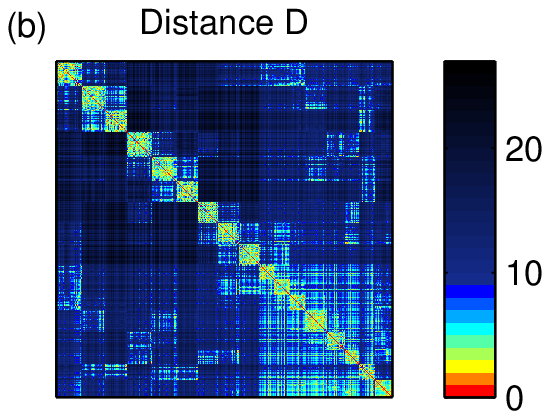} & \hspace{-0.7cm}\includegraphics{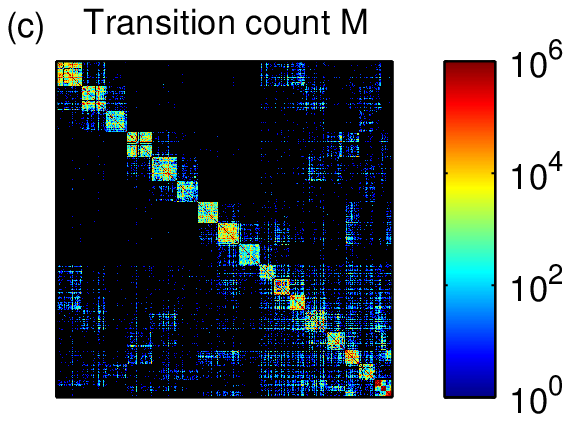} \\[-0.5cm]
      \hspace{-1cm}\includegraphics[width=0.32\textwidth]{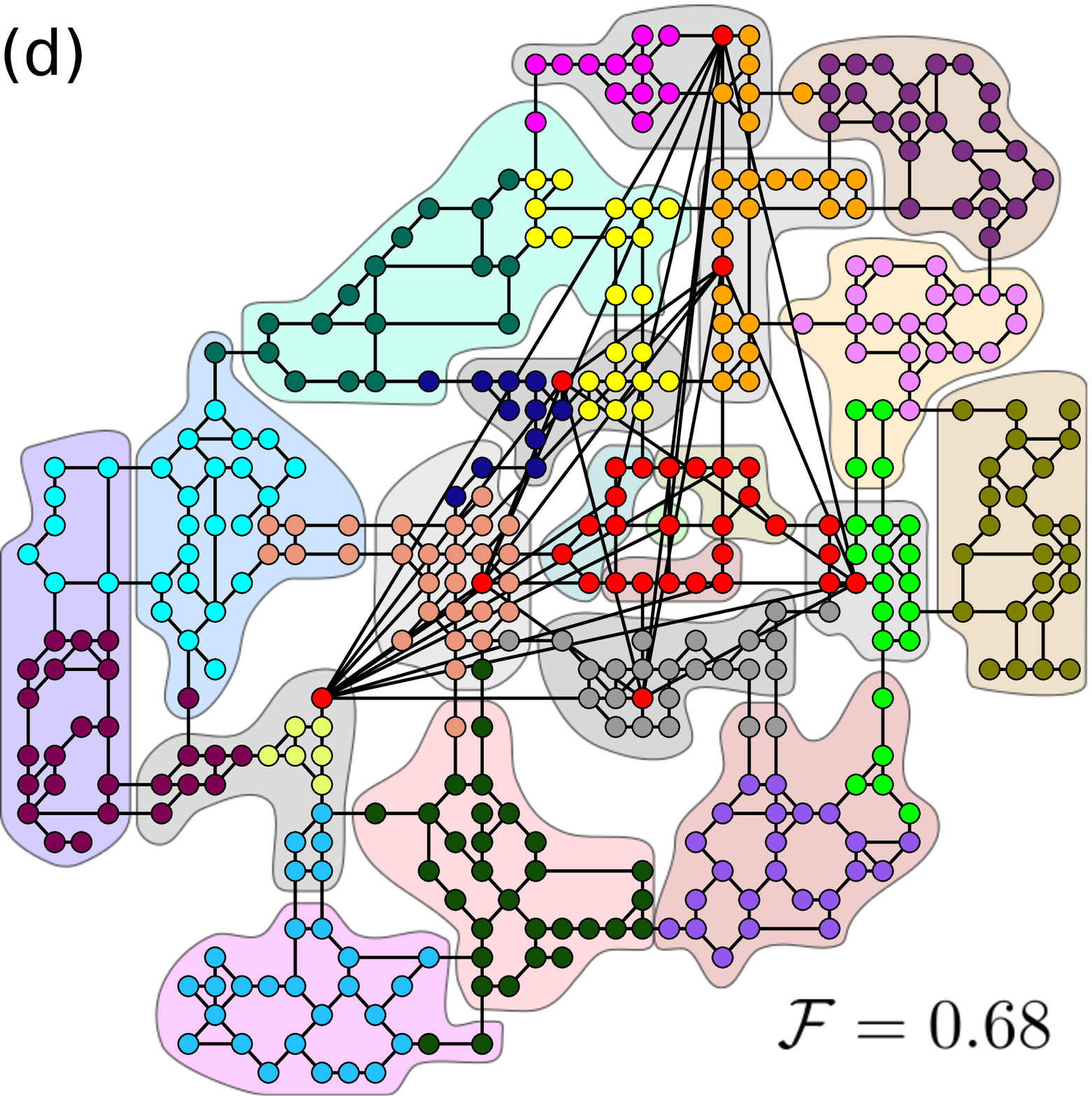} & \hspace{-1cm}\includegraphics[width=0.32\textwidth]{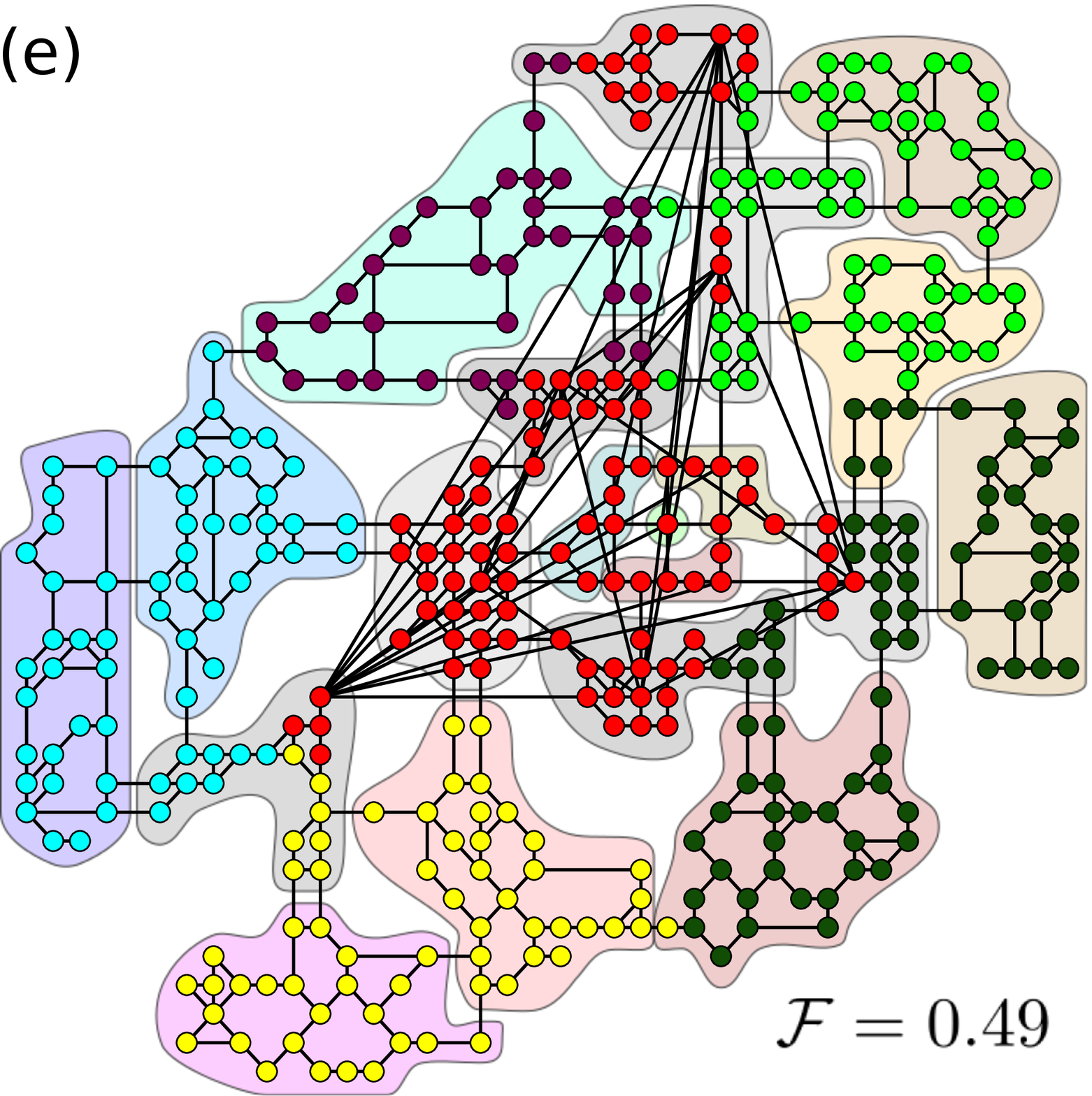} & \hspace{-1cm}\includegraphics[width=0.32\textwidth]{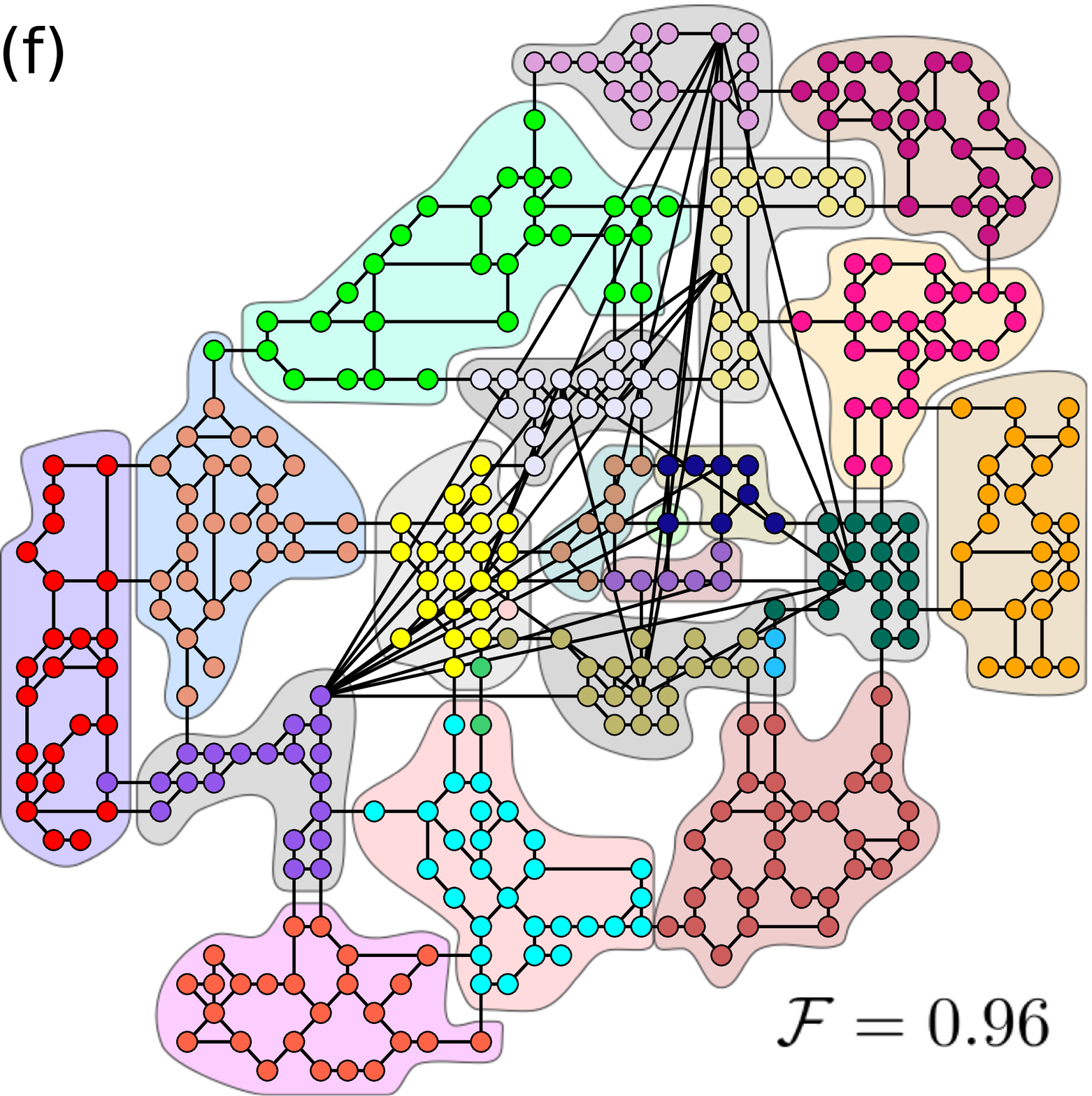} 
    \end{tabular}   
  \end{center}
  \caption{\footnotesize {\bf Extracting communities from network
      topology and from mobility patterns. } (a) The adjacency matrix
    $A$ of the universe network, (b) the matrix $D$ of shortest path
    distances, and (c) the matrix $M$ of transition counts of player jumps. Each of the
    three matrices contains $400 \times 400$ entries, whose values are  
    colour-coded. Sector IDs are ordered by cluster, resulting 
    in the block-diagonal form of the three matrices. We have used  
    modularity-optimization algorithms to extract community 
    structures from the information encoded in the three matrices. 
    Different node colours represent the different communities found, 
    while the 20 different colour-shaded areas indicate the predefined
    socio-economic clusters as in Fig.~\ref{fig:universe}. The displayed Fowlkes and Mallows index $\mathcal{F} \in \left[0,1\right]$ quantifies the overlap of the detected communities with the predefined clusters. The closer $\mathcal{F}$ is to $1$, the better the match, see \cite{SI} Section S4.
        (d) Although information contained in the adjacency matrix $A$ allows to 
    find 18 communities, a number close to the real number of clusters, 
    the communities extracted do not correspond to the underlying colour-shades 
    areas ($\mathcal{F} = 0.68$).  (e) Extracting communities from the distance matrix $D$ only 
    results in 6 different groups ($\mathcal{F}=0.49$). (f) The 23 communities detected using the transition count matrix $M$ reproduce 
  almost perfectly the real socio-economic clusters ($\mathcal{F}=0.96$), with only a few mismatched nodes detected as additional clusters. For more measures quantifying the match of communities, see \cite{SI} Table~II.
  \label{fig:communitydetection}}
\end{figure*}

Mobility patterns are influenced by the presence of the
socio-economic regions in the network, highlighted in colours in
Fig.~\ref{fig:universe}.  The typical situation is illustrated in
Fig.~\ref{fig:banana}~(a), with jumps within the same cluster being
preferred to jumps between sectors in different clusters. In order to
quantify this effect, we report in Fig.~\ref{fig:banana}~(b), blue circles, the
observed number of jumps of length $d$ within the same cluster,
divided by the total number of jumps of length $d$.  This ratio is a
decreasing function of the distance $d$, and reaches zero at $d=12$,
since no sectors at such distance do belong to the same cluster. As a
null model we report the fraction of sector pairs at
distance $d$ which belong to the same cluster, see red squares in the same figure.  The significant 
discrepancy between the two curves indicates that players indeed tend
to avoid crossing the borders between clusters. For example, a jump of
length $d=8$ from one sector to another sector in the same cluster is
expected only in $3\%$ of the cases, while it is observed 
in about $20\%$ of the cases.

Now, the propensity of a player to spend long time periods within the
same cluster might be simply related to the topology of the network,
as in the case of random walkers whose motions are constrained on graphs with strong
community structures \cite{fortunato2010report}. 
Nodes belonging to the same cluster are in
fact either directly connected or are at short distance from one
another. This proximity is reflected in the block-diagonal structure
of the adjacency matrix $A$ and of the distance matrix $D$,
respectively shown in Fig.~\ref{fig:communitydetection}~(a) and (b).
We have therefore checked whether the presence of the socio-economic
clusters originally introduced by the developers of the game can be
derived solely from the
structure of the
network. For this reason we adopted standard community detection
methods based on the adjacency and on the distance matrix \cite{Arenas:2008hq,newman2004awn}. 
The results, reported respectively in
Fig.~\ref{fig:communitydetection}~(d) and (e), show that detected communities deviate significantly from the clusters, implying that in our
online world the socio-economic regions cannot
be recovered merely from topological features. In comparison we considered the player transition count matrix $M$, shown in Fig.~\ref{fig:communitydetection}~(c), which displays a similar block-diagonal structure as $A$ and $D$, but with the qualitative difference that it contains \emph{dynamic} information on the system.
Figure~\ref{fig:communitydetection}~(f)
shows that community detection methods applied to the transition count matrix $M$ reveal almost perfectly all the socio-economic areas of the universe.  This finding demonstrates that mobility patterns
contain fundamental information on the socio-economic constraints
present in a social system. Therefore, a community detection algorithm applied to raw mobility information, as the one proposed here,
is able to extract the underlying socio-economic features, which are instead
invisible to methods based solely on topology. For a detailed treatment of adopted
community detection methods and measures see
\cite{SI} Section S4, \cite{SI} Table~II and \cite{SI} Figs.~4 and 5.

\subsection{A long-term memory model}

\begin{figure}[h!]
    \begin{center}
      \includegraphics{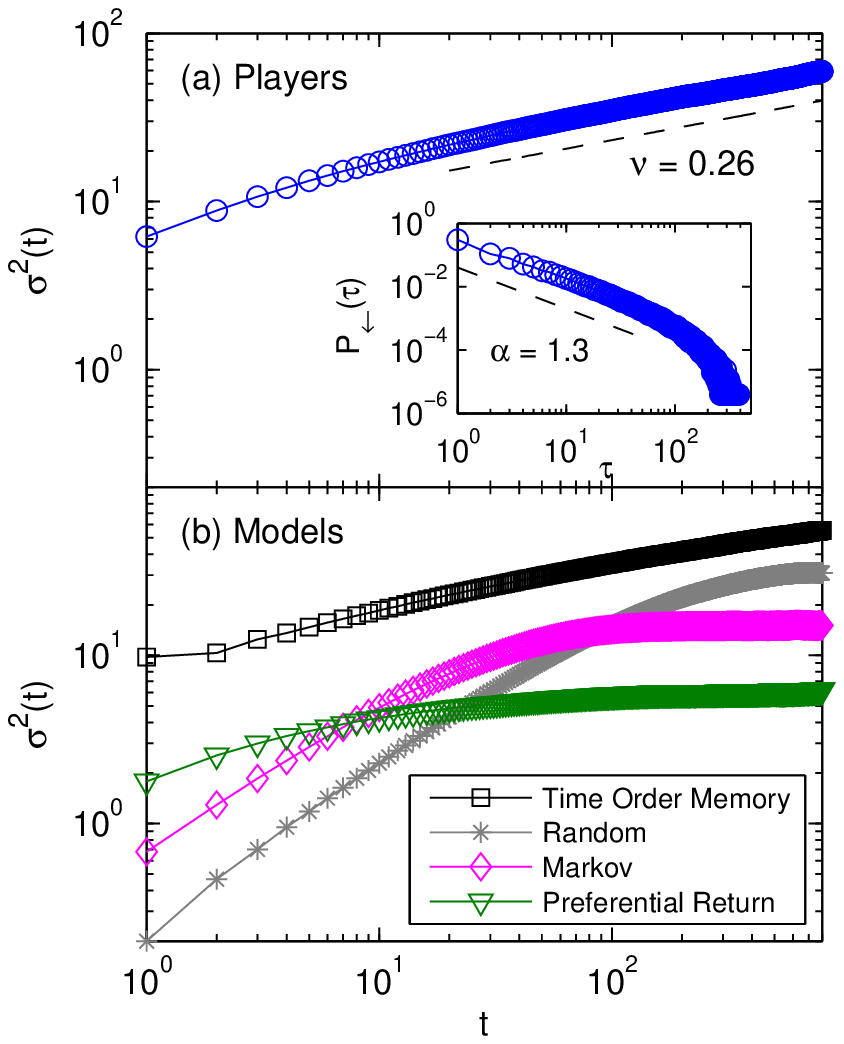}
    \end{center}
    \caption{\footnotesize {\bf Diffusion scaling in empirical data and simulated models.} (a) The mean square
      displacement (MSD) of the positions of players follows a power
      relation $\sigma^2(t) \sim t^{\nu}$ with a subdiffusive
    exponent $\nu \approx 0.26$. The inset shows the average probability
    $P_{\hookleftarrow}(\tau)$ for a player to return after $\tau$ jumps to a sector
    previously visited. The curve
    follows a power law $P_{\hookleftarrow}(\tau) \sim \tau^{-\alpha}$ with an exponent of $\alpha \approx 1.3$ and an exponential cutoff.  We report, for comparison, (b) the MSD for various models of
    mobility. For random walkers and in the case of a Markov model
    with transition probability $\pi_{ij}= m_{ij} / \sum_j m_{ij}$ we
    observe an initial diffusion with an exponent $\nu \approx 1$ and
    then a rapid saturation of $\sigma^2(t)$, due to the finite size
    of the network. A preferential return model also shows saturation
    and does not fit the empirical observed scaling exponent $\nu$.
    Conversely, a model with long-time memory (Time Order Memory) reproduces the
    exponent almost perfectly.
    Such a model makes use of the empirically observed $P_{\hookleftarrow}(\tau)$ 
    while the Markov model and the preferential return model
    over-emphasize preferences to locations visited long ago and do
    not recreate the empirical curve well. Curves are shifted vertically for visual clarity.\label{fig:msdptau}}
\end{figure}

In order to characterize the diffusion of players over the network, we
have computed the mean square displacement (MSD) of their positions,
$\sigma^2(t)$, as a function of time. Results reported in
Fig.~\ref{fig:msdptau}~(a) indicate that, for long times, the MSD
increases as a power-law:
\begin{equation}
	\sigma^2(t) \sim t^{\nu}
\label{eq:msd}
\end{equation}
with an exponent $\nu \approx 0.26$.
This anomalous subdiffusive behaviour is not a simple effect of the
topology of the Pardus universe. In fact, as shown in
Fig.~\ref{fig:msdptau}~(b), gray stars, the simulation of plain random
walks on the same network produces a standard diffusion with an
exponent $\nu \approx 1$ up to $t \approx 100$ days, and then a rapid
saturation effect which is not present in the case of the human players.

Insights from the previous section suggest that the anomalous
diffusion behaviour might be related to the tendency of players to
avoid crossing borders. We have therefore considered a
Markov model in which each walker moves from a current node $i$ to a
node $j$ with a transition probability $\pi_{ij}= m_{ij} / \sum_l m_{il}$, where $m_{ij}$ is the
number of jumps between sector $i$ and sector $j$, as expressed by the
transition count matrix $M$ of Fig.~\ref{fig:communitydetection}~(c). The 
probabilities $\pi_{ij}$ are the entries of the transition
probability matrix $\Pi$, which
contains all the information on the day-to-day movement of
real players, such as the preference to move within clusters, the
length distribution of jumps, as well as the tendency to remain in the
same sector. Despite this detailed amount of information used (the matrix $\Pi$ has 160,000 elements), the
Markov model fails to reproduce the asymptotic behaviour of the
MSD, see magenta diamonds in Fig.~\ref{fig:msdptau}~(b). Since the model considers only the position of the individual at its current time 
to determine its position at the following time, deviations from 
empirical data appear presumably due to the presence of higher-order memory effects \cite{sinatra2010nms}. 
For this reason we have considered the recently proposed preferential return model \cite{song2010msp} which incorporates a strong memory feature. The model is based on a
reinforcement mechanism which takes into account the propensity of individuals
 to return to locations they visited frequently before.
This mechanism is able to reproduce the observed tendency of
individuals to spend most of their time in a small number of
locations, a tendency which is also prevalent in the mobility behaviour
of Pardus players (see \cite{SI} Fig.~3). However, the
implementation of the preferential return model on the Pardus universe
network is not able to capture the scaling patterns of the MSD, as shown
in Fig.~\ref{fig:msdptau}~(b). The reason is that in the model the probability 
for an individual to move to a given location does not depend on the 
current location, nor on the order of previously visited locations.
Instead, we observe that in reality individuals tend to return with higher
probability to sectors they have visited recently and with lower
probability to sectors visited a long time before. Consequently a sector
that has been visited many times but with the most recent visit dating back one year
has a lower probability to be visited again than a sector that has been visited just a few times
but with the last visit dating back only one week.

To highlight this mechanism we measured the return time distribution
in the jump-time series (see Methods). In particular, we extracted the probability $P_{\hookleftarrow}(\tau)$ for an individual to return again (for the first time) to
the currently occupied sector after $\tau$ jumps. As shown in the inset of Fig.~\ref{fig:msdptau}~(a), we found that
the return time distribution reads
\begin{equation}
P_{\hookleftarrow}(\tau) \sim \tau^{-\alpha}
\label{eq:ptau}
\end{equation}
with an exponent $\alpha \approx 1.3$.
We used this information for constructing a model which takes into account 
the higher re-visiting probability of recently explored locations. In this way we can capture the long-term scaling properties of movements. Exactly these asymptotic properties are fundamentally relevant for issues of epidemics spreading or traffic management.
 
This ``Time Order Memory'' (TOM) model incorporates a power-law distribution of first return times, together with a power-law distribution of waiting times and an exponential distribution of jump distances, as those observed empirically in Fig.~\ref{fig:distributions}. We show below that these ingredients are sufficient to reproduce the subdiffusive behaviour reported in Fig.~\ref{fig:msdptau}~(a). 
The model works as follows: an individual stands still in a given sector for a
number of days drawn from the waiting time distribution,
Eq.~(\ref{eq:waitingtimedistribution}). Then, the individual jumps. There are two
possibilities: (i) with a probability $v$ she returns to an already
visited sector, (ii) with the probability $1-v$ she jumps to a so far
unexplored sector. In case (i), one of the previously visited sectors is chosen 
according to Eq.~(\ref{eq:ptau}). In the exploration case (ii), the
individual draws a distance $d$ from the distance distribution,
Eq.~(\ref{eq:distancedistribution}), and jumps to a randomly selected,
unexplored sector at that distance. 
The model has four parameters. The parameters $\lambda$, $\beta$ and $\alpha$ of equations
(\ref{eq:distancedistribution}), (\ref{eq:waitingtimedistribution}) and (\ref{eq:ptau}) respectively, are fixed by the data. Further, averaging over all jumps and players, the probability of returning to an already visited location is $v \approx 0.83$. 
Similarly to the measured data, the MSD of the TOM model, black squares in Fig.~\ref{fig:msdptau}~(b), exhibits no saturation effects and displays an exponent $\nu_{\mathrm{TOM}} = 0.23 \pm 0.02$ (the error is calculated over an ensemble of realizations) in agreement with the exponent observed for the players.

The flat slope of $\nu \approx 0.26$ and the lack of saturation of the MSD of the players over the whole observation period exposes the significant level of subdiffusivity in the motions of individuals, consistent with previous findings \cite{west1997fdl, song2010msp, metzler2000rwg, scafetta2002lsc, viswanathan1999osr}.
However, the mere tendency of individuals to return to already visited locations is not sufficient to capture these subdiffusive properties of the MSD, but it is fundamental to consider a mechanism that takes into account the temporal order of visited locations, as achieved by the TOM model.
Moreover, the TOM model is realistic in the sense that, in contrast to Markov models, it takes into account the tendency of individuals to develop a preference for visiting certain locations. At the same time it allows for the possibility that a previously preferred location becomes not frequented anymore. This view provides an alternative to recently suggested reinforcement mechanisms in preferential return models \cite{song2010msp}. The possibility for individuals to ``change home'' is relevant when the model should be able to account for migration, which is an important feature in the long-time mobility behaviour of humans. 

Finally, we discuss to which extent the findings from our ``social petri dish'' are valid also for human populations unrelated to the game. Previous analyses of human social behaviour in Pardus \cite{szell2010msd,szell2010mol} have shown agreement with well-known sociological theories and with properties on comparable behavioural data. Examining the preference of players to move \emph{within} socio-economic regions is of obvious importance for clearing up the role of political or socio-economic borders on the movement and migration of humans, where the presence of borders has a strong influence on mobility \cite{thiemann2010sbs, ratti2010rmg, newman2006lcs, lambiotte2008gdm}.
Online societies as the one of Pardus have the evident potential to serve as ``socio-economic laboratories'', where the complete knowledge of activities, social relations, and positions of all individuals can significantly advance our understanding of large-scale human behaviour, in particular of mobility.

\footnotesize
\subsection{Methods}
{\bf Data set.} We focus on one of the three \emph{Pardus} universes, \emph{Artemis}. For this universe, we extract player mobility data from day 200 to day 1200 of its existence. We discard the first 200 days because social networks between players of \emph{Pardus} have shown aging effects in the beginning of the universe, i.e. there seems to exist a transient phase in the development of the society, possibly affecting mobility, which we would like to avoid considering \cite{szell2010msd}. To make sure we only consider active players, we select all who exist in the game between the days 200 and 1200, yielding 1458 players active over a time-period of 1000 days. The sector IDs of these players, i.e. their positions on the universe network's nodes, are logged every day at 05:35 GMT. Players typically log in once a day and perform all their limited movements of the day within a few minutes, see \cite{SI} Section S1. The legal department of the Medical University of Vienna has attested the innocuousness of the used anonymized data.

{\bf Transition count matrix and transition probability matrix.} 
The entry $m_{ij}$ of the transition count matrix $M$ is equal to the number
of times a player's position was on sector $i$ and then, on the following day, on sector $j$. This number is 
cumulated for all players. The entry $\pi_{ij}$ of the transition probability matrix $\Pi$ corresponds to the probability that 
a player moves to a sector $j$ given that on the previous day the player's location was sector $i$. It reads:
$\pi_{ij}=\frac{m_{ij}}{\sum_{l}m_{il}}$, where $m_{ij}$ is the number
of observed player movements from sector $i$ to sector $j$, and the sum
over $l$ is over all sectors of the universe. The matrix $\Pi$ is a stochastic matrix, i.e. it has the property that the entries of each row sum to one. 

{\bf MSD and diffusion.} The MSD is defined as $\sigma^{2}\left(t\right)=\langle \left( r \left( T+t \right) - r\left(T\right) \right)^{2}\rangle$, where $r \left( T \right)$ and $r \left( T+t \right)$ are the sectors a player occupies at times $T$ and $T+t$ respectively, and where by $\left( r \left( T+t \right) - r\left(T\right) \right)$ we denote the distance between the two sectors. The average $\langle \cdot \rangle$ is performed over all windows of size $t$, with their left boundaries going from T=0 to T=1000-t, and over all the 1458 players in the data set.
If $\sigma^2$ has the form $\sigma^2(t) \sim t^{\nu}$ with an exponent $\nu < 1$, the diffusion process is subdiffusive, in the case $\nu>1$ it is super-diffusive. An exponent of $\nu = 1$ corresponds to classical brownian motion \cite{metzler2000rwg, west1997fdl}.

{\bf Jump-time and first return time distribution.} We transform the time-series of daily sector IDs occupied by the players from real-time to jump-time, in order to be able to compare time-series of different length and to focus on the movements between sectors. An example of this conversion is provided: a time series $\left[5, 5, 5, 32, 32, 104, 5, 5, 104, 104, 104, 32, 337, 337, 32\mathrm{\ldots}\right]$ becomes in jump-time $\left[5,32,104,5,104,32,337,32,\mathrm{\ldots}\right]$. We denote jump-time by the greek letter $\tau$, that is, at jump-time $\tau$ a player has performed exactly $\tau$ jumps. We use  $\tau$ in the computation of the first return time distribution. In the hypothetical time series of sectors $\left[5,32,104,5,104,32,337,32\right]$ a first return to a sector lying $\tau
= 1$ jumps back happens 2 times ($104,5,104$ and $32,337,32$), for $\tau = 2$
this happens once ($5,32,104,5$), for $\tau = 3$ also once ($32,104,5,104,32$). Hence, in this example, $P_{\hookleftarrow}(1) = 0.5$,
$P_{\hookleftarrow}(2) = P_{\hookleftarrow}(3) = 0.25$, where the sum over all $P_{\hookleftarrow}(\tau)$ is equal to 1. 

\subsection{Acknowledgments}
This work was conducted under the HPC-EUROPA2 project (project number: 228398) with the support of the European Commission -- Capacities Area -- Research Infrastructures initiative, and within the framework of European Cooperation in Science and Technology Action MP0801 Physics of Competition and Conflicts. M.S. and S.T. acknowledge support from the Austrian Science Fund Fonds zur F\"orderung der wissenschaftlichen Forschung P 23378, and from project EU FP7 -- INSITE. M.S., R.S. and G.P. also thank the Santa Fe Institute for the opportunities offered during the Complex Systems Summer School 2010, where this project originated.


\end{document}